\documentclass[twoside,fleqn]{article}

\usepackage{epsf,espcrc2}
\usepackage[hyperindex]{hyperref}
\newcommand\eprint[1]{\publishedeprint{#1}{#1}}
\newcommand\publishedeprint[2]{\href{http://xxx.lanl.gov/abs/#2}{#1}}
\ifx\href\undefined\newcommand\href[2]{#1}\fi
\newcommand{\ewxy}[2]{\setlength{\epsfxsize}{#2}\epsfbox{#1}}
\newcommand{\beq}{\begin{equation}}
\newcommand{\eeq}{\end{equation}}
\newcommand{\beqa}{\begin{eqnarray}}
\newcommand{\eeqa}{\end{eqnarray}}
\newcommand{\mnod}{\ensuremath{{\mQ^0}}}
\newcommand{\mpert}{\ensuremath{{\mQ^{pert}}}}
\newcommand{\mHH}{\ensuremath{{\mQ^{HH}}}}
\newcommand{\mQ}{\ensuremath{m_Q}}
\newcommand\0{\hphantom{0}}
\newcommand\ds{\displaystyle}

\title{$B$ and $B_c$ mesons with NRQCD and Clover actions}

\author{Presented by A. Ali Khan\address{Department of Physics and
  Astronomy, University of Glasgow, Glasgow, G12 8QQ, Scotland.} and
  T.~Bhattacharya\address{T-8 Group, MS B285, Los Alamos National
  Laboratory, Los Alamos, New Mexico 87545, U.S.A.}
\thanks{In collaboration with
S.~Collins and C.~T.~H. Davies, (Glasgow Univ., UKQCD Collab.),
R.~Gupta (LANL), J.~Shigemitsu (Ohio State Univ.) and J.~Sloan
(SCRI).}}

\begin{document}

\begin{abstract}
We present preliminary results from our study of the heavy-light
spectrum and decay constants. For the heavy quark, we use NRQCD at
various masses around and above the $b$ quark mass.  For the first
time, the heavy quark action and the heavy-light current consistently
include corrections at second order in the non-relativistic expansion,
as well as the leading finite $a$ corrections.  The light quarks are
simulated using a tadpole-improved Clover action at various masses in
the strange and $c$ quark region.
\end{abstract}

\maketitle

\section{INTRODUCTION}
Heavy-light hadrons are of topical interest in theoretical and
experimental particle physics. The $B$ meson decay constant $f_B$ is
required, together with the bag parameter $B_B$, for a determination
of the poorly known elements of the CKM matrix, in particular, of the
phase leading to CP-violation in nature. Comparison with experimental
results on the heavy-light spectrum available currently and in the
near future provides an important test of the reliablity of the
approximations used in these calculations.


Using NRQCD, it is possible to simulate the $b$ quark directly,
without having to use extrapolations in the heavy quark mass. Earlier
results for $f\sqrt{M}$, obtained using NRQCD
consistent through $O(1/\mnod)$ where $\mnod$ is the bare heavy quark
mass (see e.g.~\cite{Melb95,sara}), showed that $f\sqrt{M}$ has
a large slope as a function of $1/M$. Such a behavior would be 
an indication that higher order terms are important. We address this 
issue here.\looseness-1

\section{THIS SIMULATION}

In this study, the heavy quark action
and heavy-light currents are consistently improved through
$O(1/(\mnod)^2)$ at tree level.
We also include the leading $O(a^2)$ lattice spacing correction
to the lattice laplacian, and a term to compensate for the leading
$O(a)$ error in the discretization of the temporal
derivative. Furthermore, we include the leading relativistic
correction, which is expected to be the largest 
$O(1/(\mnod)^3)$ correction. This action is also consistent through
$O(v^4)$ in the heavy quark velocity in heavy-heavy systems.

We calculate the  heavy quark propagator $G_t$ on a timeslice $t$ 
using the explicitly time-reversal invariant evolution equation:
\beqa
G_{t+1} ={}&\ds \left(1-\frac{\delta^- H}{2}\right)
             \left(1-\frac{H_0}{2n}\right)^n \nonumber \\ 
           &\ds U_4^{\dagger}
             \left(1-\frac{H_0}{2n}\right)^n
             \left(1-\frac{\delta^+ H}{2}\right) G_t.
\eeqa
The free heavy quark Hamiltonian $H_0$
contains only the kinetic energy term
\beq
H_0 = -\frac{\Delta^{(2)}}{2\mnod},
\eeq
and the other terms are similar to~\cite{upsilon}:
\beqa
\delta^{\pm} H\ = \hskip-5pt
                  &\ds -\frac{g\vec{\sigma}\cdot\vec{B}}{2\mnod}
                       +\frac{ig}{8(\mnod)^2}
                          \left(\vec\Delta\cdot\vec{E^{\pm}} -
                          \vec{E^{\pm}}\cdot\vec\Delta\right) \nonumber \\
                  &\ds{} -\frac{g}{8(\mnod)^2}\vec{\sigma}\times
                          \left(\vec\Delta\times\vec{E^{\pm}} -
                          \vec{E^{\pm}}\times\vec\Delta\right) \nonumber \\
                  &\ds{} -\frac{(\Delta^{(2)})^2}{8(\mnod)^3} + 
                        \frac{a^2\Delta^{(4)}}{24\mnod} -
                        \frac{a(\Delta^{(2)})^2}{16n(\mnod)^2}.
\label{eqn:ham}
\eeqa
Here $\vec\Delta$, $\Delta^{(2)}$ and $\Delta^{(4)}$ denote the
tadpole-improved symmetric lattice discretization of the spatial
covariant derivative, the laplacian and the lattice-artifact fourth
derivative described in~\cite{upsilon}. The temporal gauge link, the
magnetic field $B$ and the electric field $E$ are also tadpole-improved
(i.e. each link is divided by $u_0 = 0.878$). $E^+$ and $E^-$ denote
the traceless forward and backward 2-leaf clover electric fields,
respectively. 

The heavy light currents ($J$) are also improved to the same order:
\beq
J = J_0 + \frac{1}{2\mnod}J_1 +\frac{1}{8(\mnod)^2} J_2.
\eeq
where, at tree level, 
\beqa
J_0 = \bar{q}\Gamma Q,\qquad
J_1 = \bar{q}\Gamma\left(\vec{\gamma}\cdot\vec{\Delta}\right) Q,
\nonumber\\
J_2 = \bar{q}\Gamma
            \left(\Delta^{(2)} + \vec{\sigma}\cdot \vec{B} -
                     2i\gamma_0 \vec{\gamma}\cdot\vec{E}\right)Q,
\label{eq:currents}
\eeqa
$q$ being the light spinor and $Q$ the nonrelativistic heavy quark
spinor whose upper two components are zero. For axial vector
currents, $\Gamma = \gamma_5\gamma_0$, and for vector currents,
$\Gamma = \gamma_\mu$.  Here $E$ denotes the  standard  4-leaf 
clover electric field. 

In addition, we have calculated the matrix elements of the other
operators with the right lattice symmetry and up to the mass dimension
of $J_2$. These are expected to mix under renormalization with the
tree level operators. As the renormalization constants (and hence the
mixing terms) are currently unknown, we do not discuss them any
further.  We simulate heavy quarks with masses $a\mnod =
1.6$,$2.0$,$2.7$ with $n=2$; $4.0$,$7.0$,$10.0$ with $n=1$, and in the
static approximation. The three lower mass values are in the region of
the $b$ quark, and the others are used to study the extrapolation of
the decay constant to the static limit.  The preliminary results
presented here are based on an ensemble of 57 quenched Wilson
configurations, fixed to Landau and then to Coulomb gauge, on a
$16^3\times 48$ lattice at $\beta = 6/g^2 = 6.0$. To increase
statistics we also include the time reversed versions of these
configurations.  For light and charm quarks we use a tadpole improved
clover action. We use two $\kappa$ values (0.119 and 0.126) around the
charm region and three (0.1369, 0.1375 and 0.13808) around the strange
quark mass.\looseness=-1

The heavy quark smearing functions are hydrogen-like wave functions
with zero or one node, or delta functions. All combinations of those
smearing functions at source and sink are used. The light quark
propagators are always smeared at the source with a Gaussian smearing
function with zero or one node, whereas at the sink we use either a
smearing identical to that at the source, or a delta function. The
results presented here are obtained with the zero-node smearing of the
light propagator at the source.

\section{THE SPECTRUM}
\subsection{Light spectroscopy}
In Table~\ref{tab:mpi}, we
list the masses of pseudoscalar and vector mesons with degenerate
clover quarks.  These masses are determined from the fall-off of the
zero-momentum meson correlator and thus for the heaviest quarks, where
one expects significant discretization effects in the dispersion
relation, they should only be taken as rough estimates. Using the
method described in~\cite{rajan}, we find the critical $\kappa$ to be
$\kappa_c = 0.13924(4)$ (in agreement with the UKQCD
estimate~\cite{richard}), the value of the isospin averaged light
quark kappa $\kappa_l = 0.13917(4)$, and the strange $\kappa$ to lie
between 0.1372(4) (determined from $K^\ast$ or $\Phi$ mesons), and
0.13755(19) (from $K$ mesons).  The inverse lattice
spacing from $m_\rho$ is $a^{-1} = 1.9(1)$ GeV.

\begin{table}
\caption{Pion masses with degenerate clover quarks.}
\label{tab:mpi}
\setlength{\tabcolsep}{0.1pc}
\begin{tabular}{|l|l|l|l|l|l|}
\hline
$\kappa$    & 0.119 & 0.126 & 0.1369 & 0.1375 & 0.13808 \\
\hline
$am_{\pi}$  &1.526(2)&1.162(2)&0.424(3)&0.363(3)&0.297(4)\\
$am_{\rho}$ &1.552(2)&1.200(2)&0.554(6)&0.51(1) &0.47(1) \\
\hline
\end{tabular}
\vskip -0.2in
\end{table}

\subsection{Quarkonia}
For each of the heavy quark masses used for the heavy-light
simulation, we also calculate the S and P wave states of the
corresponding heavy-heavy mesons, which is computationally quite
inexpensive. Results will be presented elsewhere.

\subsection{Heavy-Light: $B$ and $B_c$}

For the $B$ and $B_c$ spectrum, we study $^1S_0$, $^3S_1$, $^3P_0$,
$^3P_1$, $^3P_2$ and $^1P_1$ states. The data on $^1P_1$-$^3P_1$
mixing will be reported elsewhere. For the $^3P_2$ states, operators
belonging to both the $E$ and the $T_2$ representations of the cubic
group are used.

\subsubsection{Meson masses}

To calculate the mass of the heavy light $^1S_0$ mesons, we first
determine the energy difference $E_{\rm sim}(p^2)-E_{\rm sim}(0)$ by fitting
the ratio of non-zero and zero momentum meson correlators to an
exponential. This is then related to the mass $M_2$ by\looseness-1
\beq
E_{\rm sim}(p^2) - E_{\rm sim}(0) = \sqrt{p^2 + M_2^2} - M_2. \label{eq:rel}
\eeq
The values for $M_2$, determined using $p^2=1$ and $p^2=2$ (in
units of $(2\pi/16 a)^2$) with $\kappa=0.1369$, are presented in
the second and third columns of Table~\ref{tab:masses}. For
comparison, the fourth column lists results obtained assuming that the
non-relativistic dispersion relation holds for $p^2=1$, 
\beq
E_{\rm sim}(p^2) - E_{\rm sim}(0) =  \frac{p^2}{2M_2} \label{eq:non-rel} \ .
\eeq

Alternatively, the mass shift $\Delta = M_2 - E_{\rm sim}$ can be
calculated perturbatively. We use $\Delta$ calculated at lowest order in
$\alpha_s$ and $1/\mnod$~\cite{Colin}, but for the action used
in~\cite{upsilon}. $E_{\rm sim}(0)$ is the rate of
exponential fall-off of the zero momentum correlator. These $M_2$ 
are given in the fifth column of Table~\ref{tab:masses}.

Additionally, one can try to determine $\Delta$ from the degenerate
heavy-heavy $S$ state mesons using 
\beq
\Delta =\frac{1}{2}\Delta_{HH}=\frac{1}{2}(M_2 - E_{\rm sim})^{HH}.
\eeq
Determining $M_2^{HH}$ with the nonrelativistic dispersion relation 
(\ref{eq:non-rel}) for the heavy-heavy mesons, and $E_{\rm sim}^{HH}$ from
the fall-off of the heavy meson correlator, we  obtain the
last column of Table~\ref{tab:masses}.

We find that all three methods agree for the lighter masses within
errors. For the heaviest masses, the deviations are $\sim 10-20 \%$.
These are larger than the expected uncertainty of $\sim 0.1$ in
$\Delta^{pert}$ from higher orders in perturbation theory, and
indicate the presence of other systematic errors. Thus, in the
remainder of this paper, we use $M_2$ from the first method.
\begin{table}
\caption{$^1S_0$ meson masses at $\kappa = 0.1369$%
.}
\label{tab:masses}
\setlength{\tabcolsep}{0.08pc}
\begin{tabular}{|r|r|l|l|l|l|}
\hline
\multicolumn{1}{|c|}{\vbox{\boxmaxdepth=0pt\setbox0=\hbox{$a\mnod$}
                           \copy0\vskip-0.5\ht0}}&
\multicolumn{5}{c|}{$aM_2$}\\
&
\multicolumn{1}{c|}{$p^2=1$}&
\multicolumn{1}{c|}{$p^2=2$}&
\multicolumn{1}{c|}{$NR$}&
\multicolumn{1}{c|}{$\mpert$}&
\multicolumn{1}{c|}{$\mHH$}\\
\hline
1.6  & \02.16(2)  & \02.13(3)  & \02.18(2) &  2.13 & \02.20(3)  \\
2.0  & \02.55(2)  & \02.52(3)  & \02.56(3) &  2.52 & \02.60(5)  \\
2.7  & \03.23(4)  & \03.19(5)  & \03.24(4) &  3.22 & \03.33(7)  \\
4.0  & \04.49(7)  & \04.40(10) & \04.51(8) &  4.60 & \04.7(1)   \\
7.0  & 7.75(11) & \07.27(15) & \07.79(8) &  7.21 & \08.4(2)     \\
10.0 & 11.0(2)    & 10.2(3)    & 11.02(16) & 10.06 & 13.1(2)    \\
\hline
\end{tabular}
\end{table}

\subsubsection{Hyperfine splitting}

\begin{table}
\caption{Hyperfine splitting in lattice units.
} 
\label{tab:hyp}
\begin{tabular}{|l|l|l|l|}
\hline
\multicolumn{1}{|c|}{\vbox{\boxmaxdepth0pt\setbox0=\hbox{$a\mnod$}
                     \copy0\vskip-0.5\ht0}}&
\multicolumn{2}{c|}{$O(1/(\mnod)^2)$} &
\multicolumn{1}{c|}{$O(1/\mnod)$} \\
\multicolumn{1}{|c|}{} &
\multicolumn{1}{c|}{$\kappa=0.119$} &
\multicolumn{1}{c|}{$\kappa = 0.1369$} &
\multicolumn{1}{c|}{$\kappa = 0.137$} \\
\hline
1.6 & 0.0164(3)   & 0.0198(10) & \\
1.71&             &           & 0.020(1) \\
2.0 & 0.0141(3)   & 0.0165(9) & 0.017(1) \\
2.7 & 0.0116(2)   & 0.0127(7) &          \\
4.0 & 0.00865(15) & 0.0086(6) & 0.0091(8)\\
7.0 & 0.00548(12) & 0.0046(4) &          \\
8.0 &             &           & 0.0042(7) \\
10.0& 0.00865(15) & 0.0030(4) &           \\
\hline
\end{tabular}
\end{table}

\begin{figure}[tbhp] 
\centerline{\ewxy{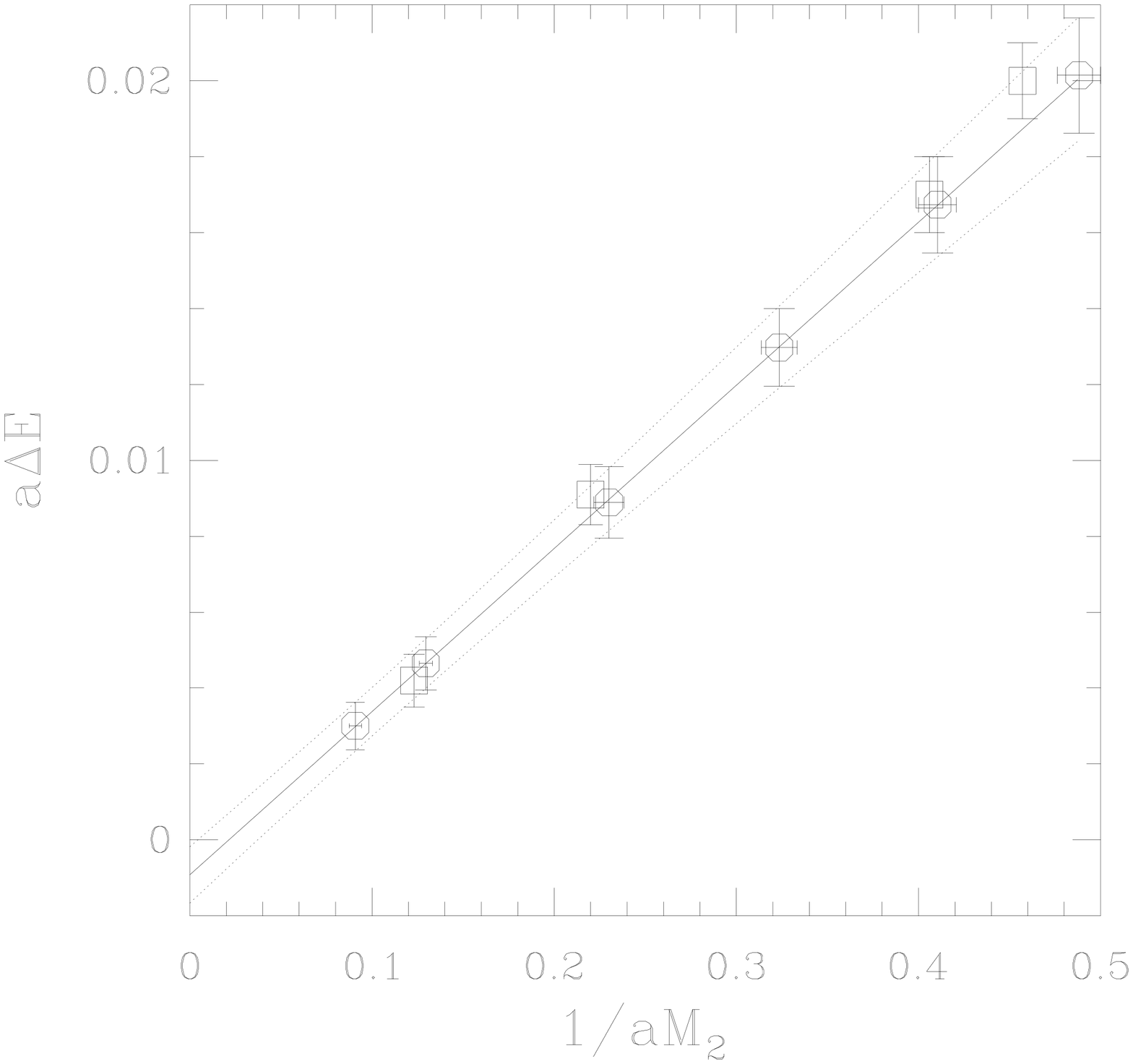}{6.8cm}}
\caption{Hyperfine splitting as a function of the mass of the $^1S_0$
    meson presented in column two of Table~\protect\ref{tab:masses},
    with a linear fit to the data. Circles are new results, whereas
    the squares are from~\protect\cite{Melb95} using an action correct
    to only $O(1/\mnod)$.}
\label{fig:hyp}
\vskip 0.3in
  \centerline{\ewxy{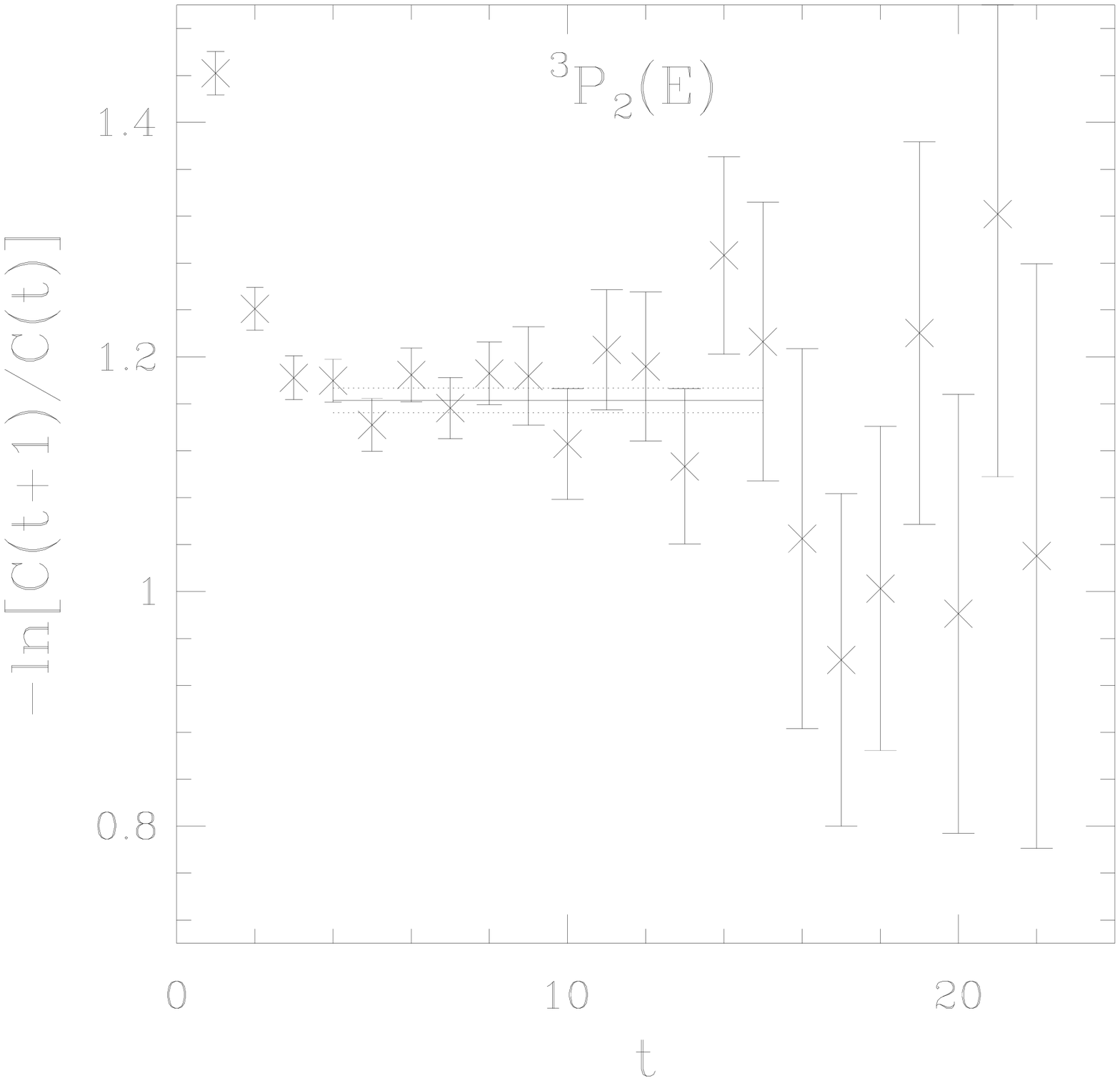}{6.8cm}}
\caption{Effective mass of a $^3P_2$ state with $\kappa =
         0.119,\;\mnod = 1.6$. Both the light and the
         heavy quarks use no-node smearing functions at the source, 
         whereas at the sink only the light quark is smeared.} 
\label{fig:3p2}
\end{figure}

We extract the hyperfine splitting $\Delta E$ directly from the
exponential fall-off of the ratio of the $^3S_1$ and the $^1S_0$
correlators. For the three light $\kappa$ values we don't observe any
variation of $\Delta E$ with $\kappa$, so we present our data without
any chiral extrapolation. Even in the charm region, the dependence of
$\Delta E$ on $\kappa$ is small. In Table~\ref{tab:hyp}, we list the
results at $\kappa = 0.119$ and $\kappa = 0.1369$, as well as the
results obtained in~\cite{Melb95} with an action correct to
$O(1/\mnod)$. For the $B$ mesons, we expect the results for the two
actions to agree, since the higher order terms in Eq.~\ref{eqn:ham} do
not affect the hyperfine splitting. The data, shown in
Fig.~\ref{fig:hyp}, substantiate this. Thus the predictions for the
hyperfine splittings in~\cite{Melb95} remain unchanged.

\subsubsection{$P$ states}

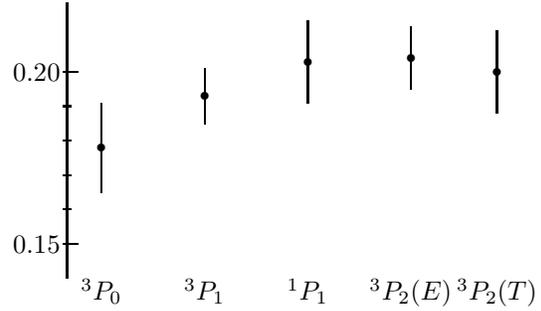
\begin{figure} 
\vspace {0.5cm}
\setlength{\unitlength}{.018in}
\begin{picture}(90,70)(0,930)
\put(10,940){\line(0,1){80}}
\multiput(9,950)(0,50){2}{\line(1,0){4}}
\multiput(9,950)(0,10){5}{\line(1,0){2}}
\put(8,950){\makebox(0,0)[r]{0.15}}
\put(8,1000){\makebox(0,0)[r]{0.20}}
\put(8,980){\makebox(0,0)[r]{ }}



\put(20,940){\makebox(0,0)[t]{$^3P_0$}}
\put(20,978){\circle*{2}}
\put(20,978){\line(0,1){13}}
\put(20,978){\line(0,-1){13}}
\put(50,940){\makebox(0,0)[t]{$^3P_1$}}
\put(50,993){\circle*{2}}
\put(50,993){\line(0,1){8}}
\put(50,993){\line(0,-1){8}}
\put(80,940){\makebox(0,0)[t]{$^1P_1$}}
\put(80,1003){\circle*{2}}
\put(80,1003){\line(0,1){12}}
\put(80,1003){\line(0,-1){12}}
\put(110,940){\makebox(0,0)[t]{$^3P_2(E)$}}
\put(110,1004){\circle*{2}}
\put(110,1004){\line(0,1){9}}
\put(110,1004){\line(0,-1){9}}
\put(135,940){\makebox(0,0)[t]{$^3P_2(T)$}}
\put(135,1000){\circle*{2}}
\put(135,1000){\line(0,1){12}}
\put(135,1000){\line(0,-1){12}}

\end{picture}
\vspace {-0.5cm}
\caption{$P-{}^1S_0$ splittings lattice units ($\kappa=0.119$, 
$\mnod=1.6$). Errors are purely statistical.}
\label{fig:Bc}
\end{figure}

A typical effective mass plot for a $P$ wave state with a heavy quark
around the $b$ mass and a light quark around the charm mass is shown
in Fig.~\ref{fig:3p2}. For lighter $\kappa$ values, the signal is only
slightly noisier. The data presented in Table~\ref{tab:pstates} and 
shown in Fig.~\ref{fig:Bc}, indicate that 
the separations between the $P$ states are qualitatively as
expected. Note, however, that as the mixing between $^1P_1$ and
$^3P_1$ has been ignored, thus the $^1P_1-{}^1S_0$ and
$^3P_1-{}^1S_0$ splittings are not individually reliable. Also, even 
though the statistical errors are of the order of the 
${}^3P_2 - {}^3P_0$  splitting, a study of the ratio of those correlators
indicates that both lattice representations of the
$^3P_2$ are equal and heavier than the $^3P_0$. 

\begin{table}
\caption{$P - {}^1S_0$ splittings for $\mnod=1.6$, $\kappa=0.119$.}
\label{tab:pstates}
\begin{center}
\begin{tabular}{|l|c|l|}
\hline
\multicolumn{1}{|c|}{state} &
\multicolumn{1}{|c|}{Lattice Rep.} &
\multicolumn{1}{|c|}{$\Delta E$} \\
\hline
$^3P_0$ & $A_1 $& 0.178(13) \\
$^1P_1$ & $T_1 $& 0.203(12) \\
$^3P_1$ & $T_1 $& 0.193(8) \\
$^3P_2$ & $E   $& 0.204(9) \\
$^3P_2$ & $T_2 $& 0.200(12) \\
\hline
\end{tabular}
\end{center}
\end{table}

\section{DECAY CONSTANTS}
\subsection{Analysis}
The decay constants are related to the one meson production amplitudes
of the local pseudoscalar or vector current $J_L$.
\beq
f\sqrt{M} = \sqrt{2} A_L=\sqrt{2/M}\langle0|J_L|B\rangle.
\eeq
To determine $A_L$, we fit to the smeared-smeared and smeared-local 
$^1S_0$ and $^3S_1$ correlators with different operator and smearing
combinations at the source and sink to the exponential form:
\beq
\langle O_{snk} O_{src} \rangle = A_{O_{src},O_{snk}}
                                      \exp (-E_{sim}^{src,snk} t).
\eeq
The measured matrix of the amplitudes is then factorized as
$A_{O_{src},O_{snk}} = A_{O_{src}} A_{O_{snk}}$ by minimizing the
overall $\chi^2$. The matrix element for the improved current,
specified in Eq.~\ref{eq:currents}, is then constructed by simply
adding the various terms.  To get the physical decay constants we
extrapolate linearly in $1/(2\kappa)$ to $\kappa=\kappa_l$ and
$\kappa=\kappa_s$. The entire procedure is carried out in a
single-elimination jackknife loop to estimate the errors in the final
quantities.

\begin{figure}[tbhp]         
\centerline{\ewxy{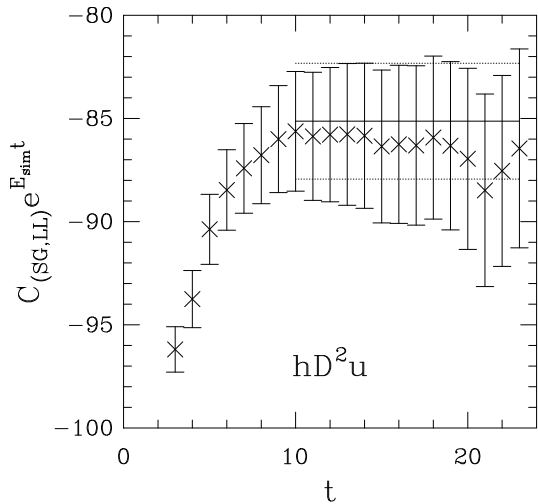}{7cm}}
\caption{Effective amplitude for the correlator of a $^1S_0$ 
         current correction at
         $\mnod=1.6$ and $\kappa=0.1369$. Both the quarks use
         smearing functions with no nodes at the source, and are
         unsmeared at the sink.}
\label{fig:amp}
\end{figure}

The signal for the effective amplitudes of the correlators show clear
plateaus except for the static case. In Fig.~\ref{fig:amp}, we show 
an example of the effective amplitude of the largest $O(1/(\mnod)^2)$
current correction to the $^1S_0$ decay constant.

\begin{table}
\caption{The contributions to the decay constants from 
         $O(1/\mnod)$ current correction (second column) and the three
         $O(1/(\mnod)^2)$ current corrections for $\kappa=0.1369$. All
         numbers in the table have been multiplied by 100 for
         convenience.}
\vskip 0.1in
\label{tab:corr}
\setlength{\tabcolsep}{0.1pc}
\begin{tabular}{|l|l|l|l|l|}
\hline
\multicolumn{1}{|c|}{$\mnod$} &
\multicolumn{1}{|c|}{$\displaystyle{\sigma\cdot D\over\strut2\mnod}$} &
\multicolumn{1}{|c|}{$\displaystyle{\sigma\cdot B\over8(\mnod)^2}$} &
\multicolumn{1}{|c|}{$\displaystyle{\sigma\cdot E\over4(\mnod)^2}$} &
\multicolumn{1}{|c|}{$\displaystyle{D^2\over8(\mnod)^2}$} \\
\hline
\multicolumn{5}{|c|}{$\kappa=0.1369$} \\
\hline
1.6   & $-5.40(19)$ & $0.82(04)$  & $1.16(05)$  & $-1.23(04)$ \\
2.0   & $-4.59(17)$ & $0.49(03)$  & $0.75(03)$  & $-0.86(03)$ \\
2.7   & $-3.59(14)$ & $0.25(02)$  & $0.41(02)$  & $-0.53(02)$ \\
4.0   & $-2.62(12)$ & $0.10(01)$  & $0.19(01)$  & $-0.27(01)$ \\
7.0   & $-1.97(07)$ & $0.035(3)$  & $0.060(5)$  & $-0.134(5)$ \\
10.0  & $-1.58(06)$ & $0.018(1)$  & $0.031(3)$  & $-0.080(3)$  \\
\hline
\multicolumn{5}{|c|}{$\kappa=0.119$} \\
\hline
1.6   & $-8.10(17)$ & $1.28(03)$  & $1.60(03)$  & $-3.27(07)$ \\
2.0   & $-7.18(16)$ & $0.78(02)$  & $1.07(02)$  & $-2.44(05)$ \\
2.7   & $-6.00(16)$ & $0.41(01)$  & $0.62(02)$  & $-1.61(04)$ \\
4.0   & $-4.67(14)$ & $0.18(01)$  & $0.30(01)$  & $-0.90(02)$ \\
7.0   & $-3.04(08)$ & $0.055(2)$  & $0.106(3)$  & $-0.39(01)$ \\
10.0  & $-2.28(07)$ & $0.025(1)$  & $0.054(2)$  & $-0.231(7)$  \\
\hline
\end{tabular}
\end{table}

\subsection{Results}

We present a study of the effect of improving the action and the
operators to $O(1/(\mnod)^2)$. It is important to recall that all
coefficients have tadpole-improved tree-level values with no
perturbative corrections. Thus, the following results
should be considered qualitative and preliminary.

\begin{figure}[tbhp]         
\centerline{\ewxy{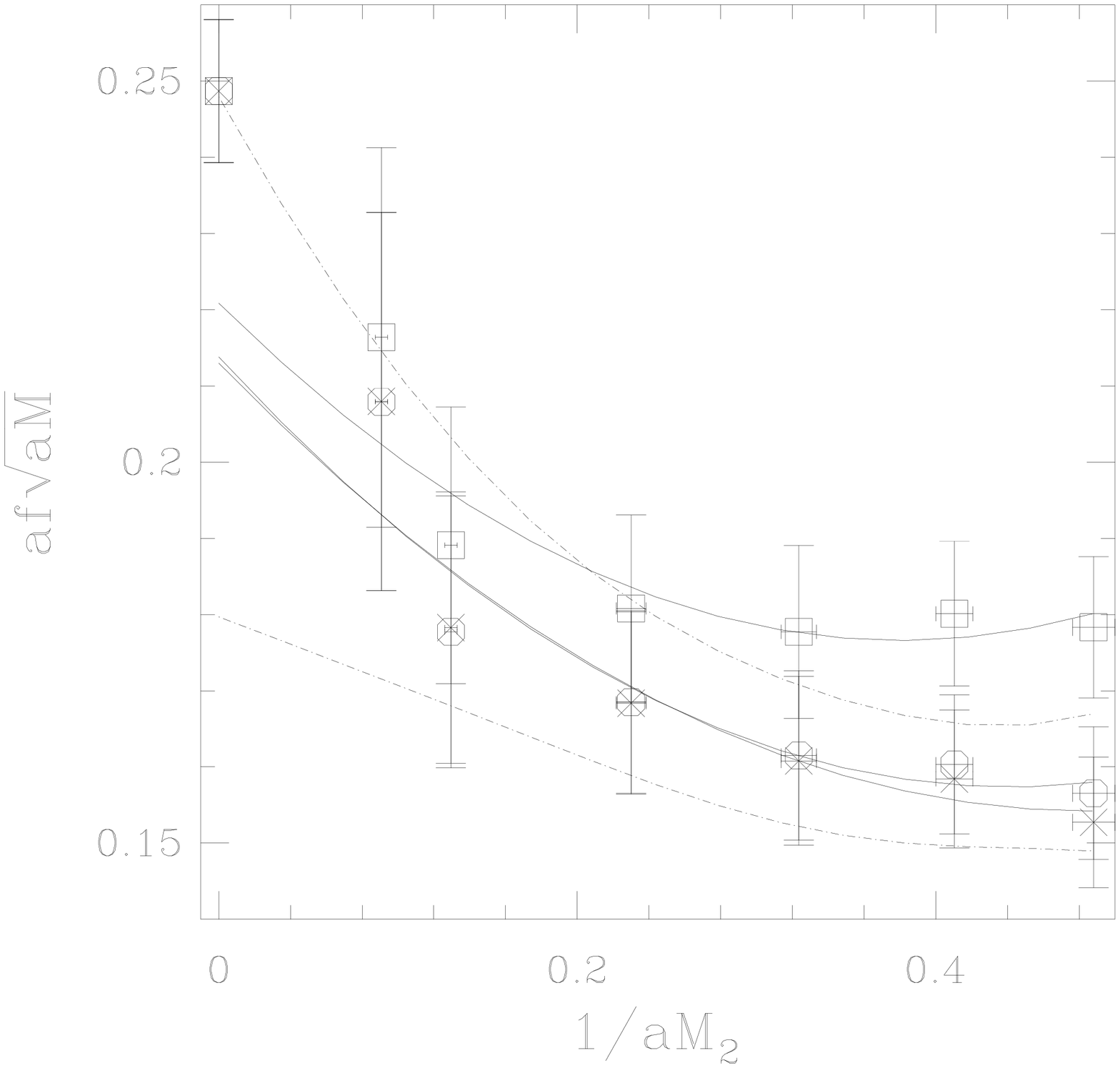}{7cm}}
\vskip -0.25in  
\caption{Comparison of the chirally extrapolated values of 
         $f \protect \sqrt{M}$ 
         with various current corrections: circles
         include all corrections through $O(1/(\mnod)^2)$,
         crosses include up to $O(1/\mnod)$ corrections, whereas
         squares are the leading order results. The fits shown are
         uncorrelated quadratic fits: broken lines show the error
         on the fit to the circles.}
\label{fig:fsqrtM}
\vspace{0.3cm}
\centerline{\ewxy{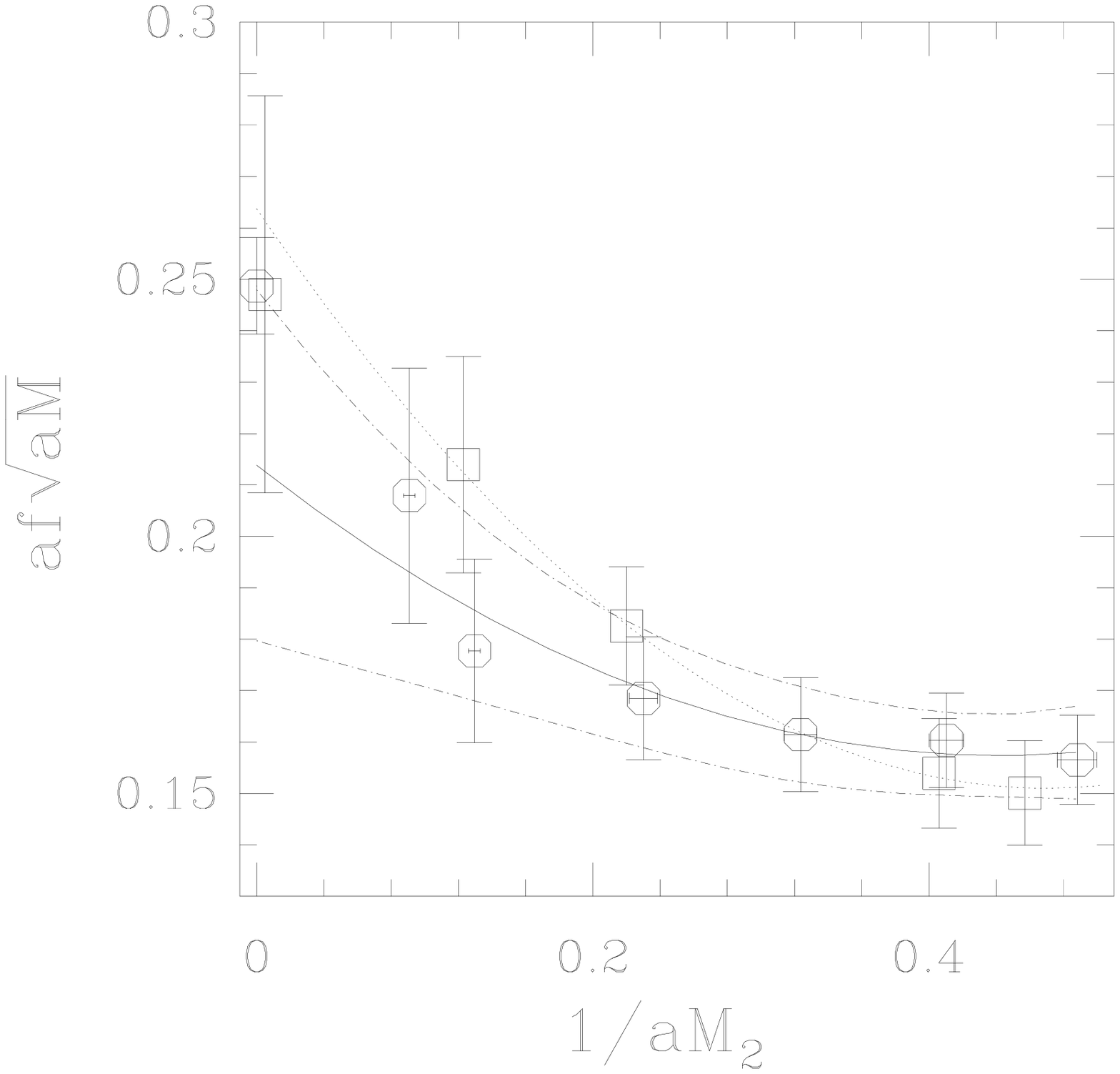}{7cm}}
\vskip -0.25in  
\caption{The chirally extrapolated $f \protect \sqrt{M}$ calculated
         consistently at $O(1/\mnod)$ from~\protect\cite{Melb95}
         (data: squares, fit: dotted line) and at $O(1/(\mnod)^2)$
         (data: circles, fit: solid line). The error in the
         $O(1/(\mnod)^2)$ points is shown by a dashed line.}
\label{fig:comp_decay}
\end{figure}

The data in Table~\ref{tab:corr} show that in the mass region of the
$B$ meson, ($\mnod\sim2.0$), the individual current corrections
arising at $O(1/(\mnod)^2)$ are $\sim 20\%$ of the $O(1/\mnod)$
corrections. However for the pseudoscalar decay constant 
the three $O(1/(\mnod)^2)$ terms add up to a result close to zero
because they have different signs.
(For the vector decay constant these three terms come with the same sign).  
The data for the different heavy quarks is shown in Fig.~\ref{fig:fsqrtM}. 

\begin{table}
\caption{$(f\protect\sqrt{M})_{PS}$ at various levels of the current
         corrections (no corrections, correct at $O(1/\mnod)$ and at
         $O(1/(\mnod)^2)$). The last column shows results of a calculation
         where both the Hamiltonian and current are correct to only 
         $O(1/\mnod)$.}
\vskip 0.1in
\label{tab:frootm}
\setlength{\tabcolsep}{0.1pc}
\begin{tabular}{|l|l|l|l|l|}
\hline
\multicolumn{1}{|c|}{$\mnod$} &
\multicolumn{1}{c|}{$O(1)$} &
\multicolumn{1}{c|}{$O((\mnod)^{-1})$} &
\multicolumn{1}{c|}{$O((\mnod)^{-2})$} &
\multicolumn{1}{c|}{\protect\cite{Melb95}} \\
\hline
1.6  & 0.178(09)  & 0.153(9)  & 0.157(9)  & \\
1.71 &            &           &           &  0.150(11) \\
2.0  & 0.180(10)  & 0.158(9)  & 0.160(9)  & 0.154(12) \\
2.7  & 0.178(11)  & 0.161(11) & 0.161(11) & \\
4.0  & 0.181(12)  & 0.169(12) & 0.168(12) & 0.182(13) \\
7.0  & 0.189(18)  & 0.178(18) & 0.178(18) & \\
8.0  &            &           &           & 0.21(2) \\
10.0 & 0.216(25)  & 0.208(24) & 0.208(25) & \\
$\infty$ & 0.249(9) & 0.249(9) & 0.249(9) & 0.25(4)\\
\hline
\end{tabular}
\end{table}

The effects of the $O(1/(\mnod)^2)$ corrections in the hamiltonian are
shown in Fig.~\ref{fig:comp_decay} where we compare the chirally
extrapolated results against the $O(1/\mnod)$ estimates given
in~\cite{Melb95}. We find that the $O(1/(\mnod)^2)$ data
indicates a smaller slope, though the difference is marginally 
significant. We are in the process of calculating the perturbative 
corrections to the coefficients in order to extract the physical value
of the decay matrix elements and their slope as a function of $1/M$.

\subsection*{Acknowledgements}
This work was supported by SHEFC, the U.S. DOE, PPARC, and the NATO
under grant number CRG 941259.  We acknowledge the ACL at LANL and
NCSA at Urbana for computational support.

\end{document}